\documentclass [prb,superscriptaddress, showpacs, twocolumn]{revtex4}
\usepackage{graphicx}
\usepackage{dcolumn}

\makeatletter

\newcommand{\Rmnum}[1]{\expandafter\@slowromancap\romannumeral #1@}
\makeatother

\begin{document}

\title{Unconventional Multiferroicity in Cupric Oxide}

\author{Pierre Tol\'{e}dano}
\affiliation{Laboratory of Physics of Complex Systems, University of Picardie, 33 rue Saint-Leu, 80000 Amiens, France}
\author{Na\"{e}mi Leo}
\affiliation{HISKP, University of Bonn, Nussallee 14-16, 53115 Bonn, Germany}
\author{Dmitry D. Khalyavin}
\affiliation{ISIS facility, STFC Rutherford Appleton Laboratory, Chilton, Didcot, Oxfordshire, OX11-0QX,United Kingdom}
\author{Laurent C. Chapon}
\affiliation{ISIS facility, STFC Rutherford Appleton Laboratory, Chilton, Didcot, Oxfordshire, OX11-0QX,United Kingdom}
\author{Tim Hoffmann}
\affiliation{HISKP, University of Bonn, Nussallee 14-16, 53115 Bonn, Germany}
\author{Dennis Meier}
\affiliation{HISKP, University of Bonn, Nussallee 14-16, 53115 Bonn, Germany}
\author{Manfred Fiebig}
\affiliation{HISKP, University of Bonn, Nussallee 14-16, 53115 Bonn, Germany}

\date{\today}

%%%%%%%%%%%%%%%%%%%%%%%%%%%%%%%%
\begin{abstract}
The magnetic phase transitions reported below 230 K in cupric oxide are analyzed theoretically at the macroscopic and microscopic levels. The incommensurate multiferroic and lock-in commensurate phases are shown to realize an inverted sequence of symmetry-breaking mechanisms with respect to the usual sequence occurring in low temperature multiferroic compounds. The higher temperature spin-spiral phase results from coupled order-parameters which decouple at the lock-in transition to the commensurate ground state phase.  Expressing the order-parameters in function of the magnetic spins allows determining the symmetries and magnetic structures of the equilibrium phases and the microscopic interactions which give rise to the polarization.
\end{abstract}


\maketitle

\begin{figure}[t]
\includegraphics[scale=1.30]{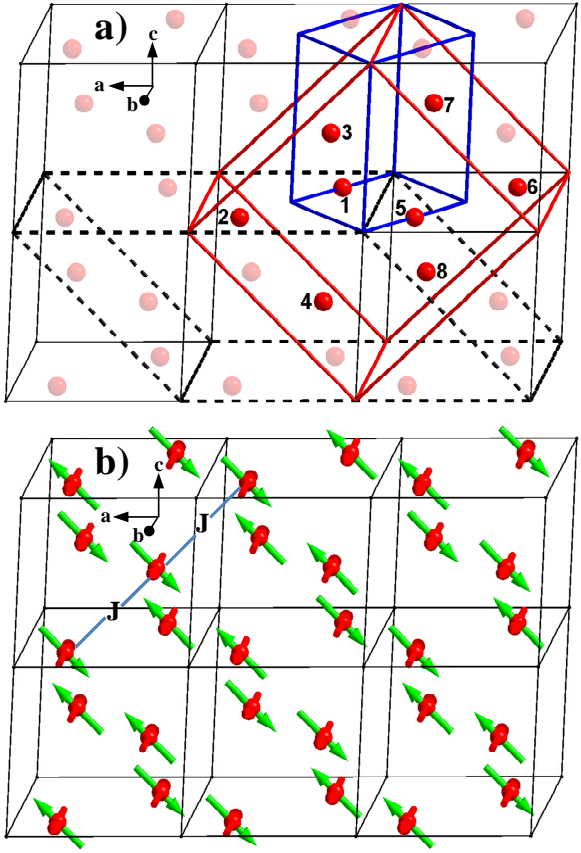}
\caption{(Color online) (a) Different unit cells used in the description of CuO. The conventional C-centred and primitive paramagnetic unit-cell are represented in black and blue solid lines, respectively. The fourfold magnetic unit cell is in solid red lines. The eight Cu atoms used in our description have the coordinates $1(\frac{1}{4},\frac{1}{4},0),2(\frac{3}{4},\frac{3}{4},0),3(\frac{1}{4},\frac{3}{4},\frac{1}{2}),4(\frac{1}{4},\frac{3}{4},-\frac{1}{2}),5(-\frac{1}{4},\frac{3}{4},0)$, $6(-\frac{3}{4},\frac{1}{4},0),7(-\frac{1}{4},\frac{1}{4},\frac{1}{2})$ and $8(-\frac{1}{4},\frac{1}{4},-\frac{1}{2})$. (b) Calculated orientation of the spins in the commensurate magnetic phase of CuO (T$<$213K) along $\vec{b} $ (red arrows) and parallel to the $(a,c)$ plane (green arrows).}
\label{fig:F1}
\end{figure}

\indent In multiferroic compounds such as TbMnO$_3$,\cite{ref:1} MnWO$_4$ \cite{ref:2} or Ni$_3$V$_2$O$_8$, \cite{ref:3} the magnetically induced ferroelectric order does not appear directly below the paramagnetic phase but across an intermediate antiferromagnetic  phase via two successive second-order phase transitions. This typical two-step phase-sequence involves two order-parameters associated with the same incommensurate wave-vector.\cite{ref:4,ref:5,ref:6} One order-parameter induces the paramagnetic-antiferromagnetic transition; it then couples to a second order-parameter producing a complex spiral-type magnetic order which breaks space inversion and gives rise to an electric polarization. Such coupling, which expresses the existence of
competing magnetic interactions, tends to lower the transition temperature to the multiferroic phase. This phase constitutes in most cases the ground state of the system.\\
\indent The preceding picture of magnetically driven multiferroics was recently blurred by the discovery of a multiferroic spin spiral phase in cupric oxide,\cite{ref:7} in which the multiferroic and antiferromagnetic phases occurs in an inverted order with respect to the standard sequence found in low temperature multiferroics. The multiferroic phase appears directly below the paramagnetic phase at $T_{N2}=230 K$. At $T_{N1}=213 K$ the incommensurate spiral ordering is replaced  by a non-polar commensurate spin structure representing the actual ground state of the material. The remarkably high transition temperature $T_{N2}$  has been attributed to large superexchange spin interactions,\cite{ref:7,ref:8} and incidentally related to the property of CuO to be a starting material for the synthesis of high-$T_C$ superconductors.\cite{ref:7} Here we describe theoretically the sequence of transitions in CuO and show that the corresponding transition mechanisms, which both display a first-order character, also occur in a reversed order: The higher temperature multiferroic phase results from the coupling of two order-parameters whereas the low-temperature commensurate phase is induced by a single order-parameter.  Expressing the order-parameter components in terms of spin densities allows to detail the magnetic structures of the phases reported from neutron diffraction measurements\cite{ref:9,ref:10,ref:11,ref:12} and to elucidate the nature of the interactions giving rise to the induced polarization.\\
\indent The wave-vector $\vec{k}_c=(\frac{1}{2},0,-\frac{1}{2})$  reported for the transition to the commensurate phase\cite{ref:9} arising at $T_{N1}$ is located \textit{inside} the monoclinic $C$ Brillouin-zone. It is invariant by the mirror-plane $m_y$ of the paramagnetic space-group $C2/c1'$, corresponding to a two-branch star $k^*_c=\pm k_c$ associated with two bi-dimensional irreducible representations (IR's) $\tau_1$ and $\tau_2$,\cite{ref:13} the matrices of which are given in Table \ref{table:I}. The Landau free-energy expressing the coupling between the corresponding 2-component order-parameters ($\eta_1=\rho_1cos\theta_1,\eta_2=\rho_1sin\theta_1$) and $(\zeta_1=\rho_2cos\theta_2, \zeta_2=\rho_2sin\theta_2)$ reads:
\begin{eqnarray}
F = \frac{\alpha_1}{2}\rho^2_1 + \frac{\beta_1}{4}\rho^4_1 + \frac{\gamma_1}{4} \rho^4_1cos4\theta_1+ \cdot \cdot \cdot \nonumber \\
+ \frac{\alpha_2}{2}\rho^2_2 + \frac{\beta_2}{4}\rho^4_2 + \frac{\gamma_2}{4} \rho^4_2cos4\theta_2+ \cdot \cdot \cdot \nonumber \\
+ \frac{\nu_1}{2}\rho^2_1\rho^2_2 cos 2(\theta_1+\theta_2) + \frac{\nu_2}{2} \rho^2_1\rho^2_2cos2(\theta_1-\theta_2) 
\label{eq:F1}
\end{eqnarray}
Minimization of $F$ truncated at the \textit{eighth} degree leads to eleven possible stable phases for different equilibrium values of $\rho_1,\rho_2,\theta_1$ and $\theta_2$. The magnetic symmetries of the phases, numbered from $\Rmnum{1}$ to $\Rmnum{11}$, are listed in Table \ref{table:II} together with the equilibrium values of the order-parameters, and the basic translations of their fourfold primitive unit-cells. In order to identify the order-parameter symmetries associated with the sequence of phases reported in CuO, let us express the order-parameter components in function of the magnetic spins. Denoting $\vec{s}_1-\vec{s}_8$ the spins associated with the eight Cu$^{2+}$ ions in the magnetic unit-cell shown in Fig. \ref{fig:F1}(a), one can write $\vec{s_i}=s^a_i\vec{a}_m+s^b_i\vec{b}_m+s^c_i\vec{c}_m$, where $\vec{a}_m=(a,0,c),\vec{b}_m=(0,b,0),\vec{c}_m=(a,0,-c)$, $a,b,c$ being the lattice parameters of the paramagnetic $C$-centred unit-cell. Projecting the representation transforming the $s^{\mu}_i(\mu =a,b,c)$ on $\tau_1$  and $\tau_2$ one finds three copies of each order-parameter $(\eta_1,\eta_2)$ and $(\zeta_1,\zeta_2)$ in function of the spin components, which are: 
\begin{eqnarray}
\eta^{ac}_1 = s^{a,c}_1-s^{a,c}_6-s^{a,c}_7+s^{a,c}_8, \nonumber \\
\eta^{a,c}_2 = -s^{a,c}_2+s^{a,c}_3-s^{a,c}_4+s^{a,c}_5; \nonumber \\
\eta^{b}_1 = s^{b}_1-s^{b}_6+s^{b}_7-s^{b}_8, \eta^{b}_2 = -s^{b}_2-s^{b}_3+s^{b}_4+s^{b}_5; \nonumber \\ 
\zeta^{ac}_1 = s^{a,c}_2+s^{a,c}_3-s^{a,c}_4-s^{a,c}_5, \nonumber \\
\zeta^{ac}_2 = s^{a,c}_1-s^{a,c}_6+s^{a,c}_7-s^{a,c}_8; \nonumber \\
\zeta^{b}_1 = s^{b}_2-s^{b}_3+s^{b}_4-s^{b}_5, \zeta^{b}_2 = s^{b}_1-s^{b}_6-s^{b}_7+s^{b}_8;
\label{eq:F2}
\end{eqnarray}
Taking into account the equilibrium conditions fulfilled by the order-parameter components, indicated in column (d) of Table \ref{table:II}, the magnetic structures associated with the eleven phases listed in Table \ref{table:II} can be worked out. Using the neutron diffraction data on CuO\cite{ref:9,ref:10,ref:11,ref:12} then allows an unambiguous identification of the phases disclosed in this compound. Thus, the commensurate phase found below $T_{N1}$ corresponds to phase $\Rmnum{4}$ induced by $\tau_2$ for $\zeta_1=-\zeta_2$. It has the magnetic symmetry $P_a2_1/c$ with the unit-cell basic vectors $(2a,0,0)$, $(0,b,0)$ and $(a,0,c)$ (Fig. \ref{fig:F1}(a)) and the anti-translations $\frac{c\pm a}{2}$. Its magnetic structure is determined by the equilibrium conditions $\eta^{a,b,c}_1=0, \eta^{a,b,c}_2=0, \zeta^{a,b,c}_1=-\zeta^{a,b,c}_2$, and the relationships imposed by the anti-translations, which are $\vec{s}_1=-\vec{s}_6,\vec{s}_2=-\vec{s}_5,\vec{s}_3=-\vec{s}_4,\vec{s}_7=-\vec{s}_8$. It yields:
\begin{eqnarray}
s^{a,c}_1=s^{a,c}_4=s^{a,c}_5=s^{a,c}_7= \nonumber \\
-s^{a,c}_2=-s^{a,c}_3=-s^{a,c}_6=-s^{a,c}_8, \nonumber \\
s^{b}_1=s^{b}_3=s^{b}_5=s^{b}_8=-s^{b}_2=-s^{b}_4=-s^{b}_6=-s^{b}_7 
\label{eq:F3}
\end{eqnarray}
\begin{table*}[t]
\caption{Irreducible representations $\tau_1$ and $\tau_2$ of the paramagnetic space group $C2/c1'$ associated with the $\vec{k}_c=\pm (\frac{1}{2},0,-\frac{1}{2})$ wave vector star. $T$ is the time reversal operator.}% title of Table
\centering % used for centering table
\begin{tabular*}{1.00\textwidth}{@{\extracolsep{\fill}} c | c c c c c c c c} % centered columns (4 columns)
\hline \hline \\
$ C2/c $ & $\lbrace 2_y\mid 0 0 \frac{c}{2} \rbrace$ & $\lbrace m _y\mid 0 0 \frac{c}{2} \rbrace$ & $\lbrace \bar{1} \mid 0 0 0 \rbrace$ & $T$ & $\lbrace 1\mid a 0 0 \rbrace$ & $\lbrace 1\mid 0 b 0 \rbrace$ & $\lbrace 1\mid 0 0 c \rbrace$ & $\lbrace 1\mid \frac{a}{2} \frac {b}{2}0 \rbrace$ \\ [1.0ex] % inserts table
%heading
\hline \\ % inserts single horizontal line
$ \tau_1 \begin{array}{c} \eta_1 \\ \eta_2 \end{array}$ & $\left [ \begin{array}{cc} 1 & 0 \\ 0 & -1 \end{array} \right ]$ & $\left [ \begin{array}{cc} 0 & 1 \\ -1 & 0 \end{array} \right ]$ & $\left [ \begin{array}{cc} 0 & 1 \\ 1 & 0 \end{array} \right ]$ & $\left [ \begin{array}{cc} -1 & 0 \\ 0 & -1 \end{array} \right ]$ & $\left [ \begin{array}{cc} -1 & 0 \\ 0 & -1 \end{array} \right ]$ & $\left [ \begin{array}{cc} 1 & 0 \\ 0 & 1 \end{array} \right ]$ & $\left [ \begin{array}{cc} -1 & 0 \\ 0 & -1 \end{array} \right ]$ & $\left [ \begin{array}{cc}  0 & 1 \\ -1 & 0 \end{array} \right ]$ \\ % inserting body of the table  
\\
\hline \\ % inserts single horizontal line
$ \tau_2 \begin{array}{c} \zeta_1 \\ \zeta_2 \end{array}$ & $\left [ \begin{array}{cc} 1 & 0 \\ 0 & -1 \end{array} \right ]$ & $\left [ \begin{array}{cc} 0 & -1 \\ 1 & 0 \end{array} \right ]$ & $\left [ \begin{array}{cc} 0 & -1 \\ -1 & 0 \end{array} \right ]$ & $\left [ \begin{array}{cc} -1 & 0 \\ 0 & -1 \end{array} \right ]$ & $\left [ \begin{array}{cc} -1 & 0 \\ 0 & -1 \end{array} \right ]$ & $\left [ \begin{array}{cc} 1 & 0 \\ 0 & 1 \end{array} \right ]$ & $\left [ \begin{array}{cc} -1 & 0 \\ 0 & -1 \end{array} \right ]$ & $\left [ \begin{array}{cc}  0 & 1 \\ -1 & 0 \end{array} \right ]$ \\
\\
\hline
\end{tabular*}
\label{table:I} 
\end{table*}
\begin{table}[b]
\caption{Magnetic space groups (Column (c)) deduced from the minimization of the free-energy  (Eq. (\ref{eq:F1}));  Column (a): irreducible or reducible representations; Column(b): Stable state number; Column (d) Equilibrium values of the order-parameters; Column (e) Basic vectors of the fourfold magnetic unit-cell.}% title of Table
\centering % used for centering table
\begin{tabular*}{0.48\textwidth}{@{\extracolsep{\fill}} c | c c c c} % centered columns (4 columns)
\hline \hline \\ % inserts single horizontal line
(a) & (b) & (c) & (d) & (e) \\[1.0ex]
\hline\\
$ \tau_1 $ & $\begin{array}{c} \Rmnum{1} \\ \Rmnum{2} \\ \Rmnum{3} \end{array}$ & $\begin{array}{c} P_a2_1/c \\ P_a2/c \\ P_ac \end{array}$ & $\begin{array}{c} \eta_1=\eta_2 \\ \eta_1=0 \\ \eta_1 \neq \eta_2 \end{array}$ & $2c, b, c-a $ \\
\\ [0.5ex] % inserts table
\hline \\ 
$ \tau_2$ & $\begin{array}{c} \Rmnum{4} \\ \Rmnum{5} \\ \Rmnum{6} \end{array}$ & $\begin{array}{c} P_a2_1/c \\ P_a2/c \\ P_ac \end{array}$ & $\begin{array}{c} \zeta_1=-\zeta_2 \\ \zeta_2=0 \\ \zeta_1 \neq \zeta_2 \end{array}$ & $2a, b, a+c $ \\
\\ [0.5ex] % inserts table
\hline \\ 
$ \tau_1 + \tau_2 $ & $\begin{array}{c} \Rmnum{7} \\ \Rmnum{8} \\ \Rmnum{9} \\ \Rmnum{10} \\ \Rmnum{11} \end{array}$ & $\begin{array}{c} P_s\bar{1} \\ P_a2_1 \\ P_s\bar{1} \\ P_a2 \\ P_s1  \end{array}$ & $\begin{array}{c} \eta_1=\eta_2, \zeta_1=-\zeta_2 \\ \eta _1 = \eta _2, \zeta_1=\zeta_2 \\ \eta_1=0,\zeta_2=0 \\ \eta_1=0, \zeta_1=0 \\ \eta_1 \neq \eta 2, \zeta_1 \neq \zeta_2 \end{array}$ & $\begin{array}{c} b,c-a,2c \\ 2c,b,c-a \\ b,c-a,2c \\ 2c,b,c-a \\ b, c-a, 2c \end{array}$\\
\\
\hline
\end{tabular*}
\label{table:II} 
\end{table}
Fig. \ref{fig:F1}(b) shows the corresponding orientations of the spins along the $\vec{b}$ axis and parallel to the $(a,c)$ plane. The antiferromagnetic configuration with spins along $\vec{b}$ has been proposed experimentally\cite{ref:9,ref:10,ref:11,ref:12} based on neutron diffraction analysis. The model reflects a strong antiferromagnetic super-exchange coupling J along the $[10\bar{1}]$ direction. In addition, our symmetry-based approach predicts that non-zero spin components are allowed in the $(a,c)$ plane. The coupling between the in-plane components is, however, ferromagnetic along $[10\bar{1}]$, pointing to their probable relativistic origin and therefore too small value to be detected in usual diffraction experiments. \\
\begin{figure}[t]
\includegraphics[scale=1.25]{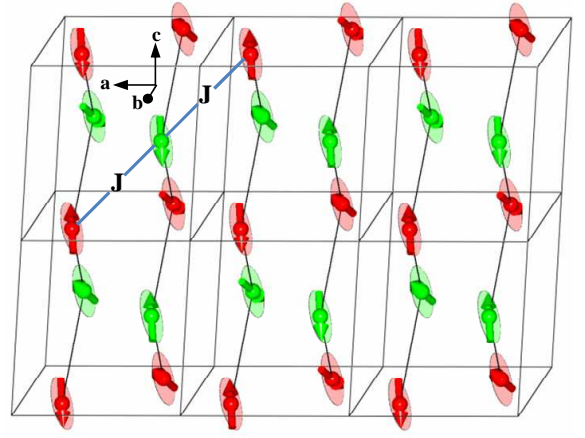}
\caption{(Color online) Calculated spin distribution in the commensurate approximant of the incommensurate structure of CuO (213 K$<$T$<$230 K). The sites related by the screw axes $2_1$(red and green) of the paramagnetic $C2/c1'$ space group are connected by solid lines.}
\label{fig:F2}
\end{figure}
\indent The polar phase $\Rmnum{8}$ of magnetic symmetry $P_a2_1$, induced by $\tau_1+\tau_2$, can be taken as a commensurate approximant of the incommensurate multiferroic phase observed below  $T_{N2}$. The corresponding equilibrium relationships, $\eta^{abc}_1=\eta^{abc}_2$ and $\zeta^{abc}_1=\zeta^{abc}_2$ give:
\begin{eqnarray}
s^{ac}_1=s^{ac}_3=-s^{ac}_4=-s^{ac}_6; s^{ac}_2=-s^{ac}_5=s^{ac}_7=-s^{ac}_8 \nonumber \\
s^{b}_1=-s^{b}_3=s^{b}_4=-s^{b}_6; -s^{b}_2=s^{b}_5=s^{b}_7=-s^{b}_8
\label{eq:F4}
\end{eqnarray} 
Figs. 2 shows the corresponding orientation of the spins compatible with the neutron diffraction analysis\cite{ref:11,ref:12} of a spiral structure having its envelope parallel to the $(b^*;0.5a^*+1.5c^*)$ plane in reciprocal space. The antiparallel orientation of the spins along the Cu-O chains run along the $[10\bar{1}]$ direction, with the strongest antiferromagnetic super-exchange coupling J. 
Note that using phase $\Rmnum{10}$ of symmetry $P_a2$ (Table \ref{table:I}) as an approximant of the spin-spiral phase of CuO, would lead to the same incommensurate magnetic structure which displays the \textit{point-group} symmetry $2_y1'$.\\ 
\begin{figure}[b]
\includegraphics[scale=1.05]{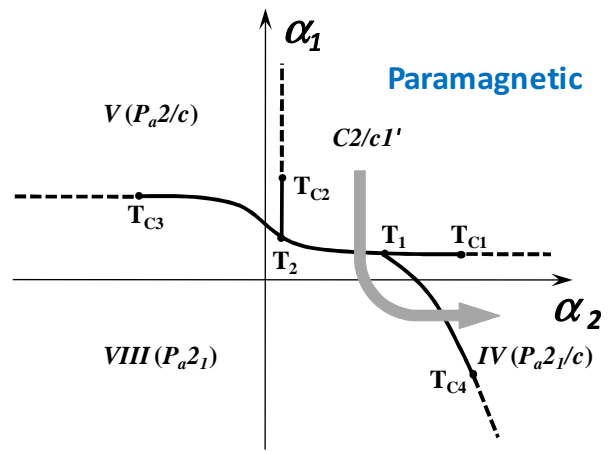}
\caption{(Color online) Theoretical phase diagram associated with the free-energy $F$, involving the commensurate phase and  approximant of the incommensurate phase of CuO. Solid and hatched curves are first and second-order transition curves. T$_{C1}$-T$_{C4}$ are tricritical points. $T_1$ and $T_2$ are triple points. The arrow shows the thermodynamic path followed on cooling in CuO.}
\label{fig:F3}
\end{figure}
\indent Thus, on decreasing temperature below $T_{N2}$ the incommensurate phase of CuO results from the coupling of two order-parameters which are decoupled at $T_{N1}$, only one of them being associated with the symmetry-breaking mechanism giving rise to the low-temperature commensurate phase. Figure \ref{fig:F3} shows the thermodynamic path corresponding to the preceding sequence of phases in the theoretical phase diagram of CuO. Note that the paramagnetic to incommensurate phase transition \textit{has necessarily a first-order character} since all the phases induced by $\tau_1+\tau_2$ in Table \ref{table:II} cannot be reached directly from the paramagnetic phase across a second-order phase transition.\cite{ref:18} \\
\indent The polar symmetry of the incommensurate phase allows an electric polarization component $P_y$, the form of which can be deduced from the dielectric free-energy $\frac{P^2_y}{2\chi^0_{yy}}-\delta(\eta_1\zeta_1+\eta_2\zeta_2)$, where $\delta$ is a coupling constant and $\chi^0_{yy}$ the paramagnetic susceptibility. It gives:
\begin{equation}
P_y=\delta \chi^0_{yy}(\eta_1\zeta _1+\eta_2\zeta_2)
\label{eq:F5}
\end{equation}
Eq. (\ref{eq:F5}) is consistent with the \textit{linear} increase of $P_y(T)$ on cooling observed below  $T_{N2}$.\cite{ref:7}  It reflects a typical \textit{improper} ferroelectric behaviour, which is confirmed by the upward finite discontinuity of the dielectric susceptibility. By contrast the absence of noticeable discontinuity at $T_{N2}$ for the polarization\cite{ref:7} does not allow confirming the first-order nature of the transition imposed by symmetry and topological requirements.\cite{ref:18} In this respect, additional precise experimental studies are necessary to verify this important point. Taking into account the order-parameter copies $P_y$ reads:
\begin{eqnarray}
P_y=\sum_{u,\nu =a,b,c}\delta_{u\nu }(\eta^u_1\zeta^{\nu }_1+\eta^u_2\zeta^{\nu }_2)
\label{eq:F6}
\end{eqnarray}
The $\delta_{ab} / \delta_{ba}$ and $\delta_{cb} / \delta_{bc}$ coupling invariants contain exclusively antisymmetric invariants of the $s^{u}_is^{\nu}_j-s^{\nu}_is^{u}_j$ type, which represent the Dzyaloshinskii-Moriya (DM) "cycloidal" interactions coupling the $s^b_i$ components to the $s^a_i$ or $s^c_i$ spin-components. The $\delta_{ac} / \delta_{ca}$ terms in Eq. (\ref{eq:F6}) express symmetric interactions of the $s^{u}_is^{\nu}_j+s^{\nu}_is^{u}_j$ type corresponding to anisotropic exchange between spins, which is usullay much weaker than DM terms.\cite{ref:21}\\
\indent At the phenomenological level the incommensurability in cupric oxide is related to the existence of anti-symmetric (Lifshitz) invariants of the form $\eta_1\frac{\partial \eta_2}{\partial u}-\eta_2\frac{\partial \eta_1}{\partial u}$ and $\zeta_1\frac{\partial \zeta_2}{\partial u}-\zeta_2\frac{\partial \zeta_1}{\partial u} (u=x,z)$ permitted by the order-parameter symmetries. The free-energy of the incommensurate phase reads: $\Phi = \int \left [ F_1(\eta_i(u),\zeta_i(u))+F_2(\eta_i,\zeta_i,\frac{\partial \eta _i}{\partial u},\frac{\partial \zeta_i}{\partial u}) \right ]dV$, where the sum runs over the volume of the system. $F_1$ is given by 
Eq. (\ref{eq:F1}) with $\gamma_i=0, \nu_i=0$, and  $F_2=\sum_{u=x,z} \lbrace \delta _u \left (\eta _1\frac{\partial \eta_2}{\partial u}-\eta_2\frac{\partial \eta_1}{\partial u} \right)+\sigma_u \left [ \left (\frac{\partial \eta _1}{\partial u}\right )^2+\left (\frac{\partial \eta_2}{\partial u}\right )^2 \right ] \rbrace +\sum_{i=1,2; u,\nu = x,z}\sigma_{u\nu}\left (\frac{\partial \eta_i}{\partial u}\frac{\partial \eta_i}{\partial \nu } \right)+(\eta_i\rightarrow \zeta_i)$. Minimization of $\Phi $ yields the incommensurate wave-vector $\vec{k}_{inc}=\vec{k}_c+(\Delta k_x,0,\Delta k_z)$ with $\Delta k_x \approx \frac{2\delta_x\sigma_z-\delta_z\sigma_x}{\sigma^2_{xz}-4\sigma_x\sigma_z}$ and $\Delta k_z=\Delta k_x(x\leftrightarrow z)$, corresponding in CuO to $\vec{k}_{inc}=(0.506,0,-0.483)$. Close below $T_{N2}$ the Lifshitz 
invariants favour the onset of the incommensurate phase. On cooling, the order-parameter magnitude increases and fourth-degree terms in $F_1$ influence the stability of the system. The first-order lock-in transition to a commensurate phase at $T_{N1}$ coincides with the onset of the anisotropic $\gamma_i$ and $\nu_i$ invariants in Eq. (\ref{eq:F1}) and with a cancelling of the Lifshitz invariants.\\
\indent In summary, our theoretical analysis of the unusual sequence of phases observed in cupric oxide shows that the corresponding transition mechanisms follows in many respects an unconventional scheme as compared to the situation found in low-temperature multiferroic compounds. The phases arise in an inverted sequence across \textit{first-order} transitions, the higher-temperature multiferroic phase, induced by the coupling of two antiferromagnetic order-parameters having a lower magnetic symmetry than the low-temperature phase associated with a single order-parameter. The order-parameters have been expressed in function of the spin variables permitting a theoretical  determination of the magnetic structures and of the magnetic interactions contributing to the electric polarization. In this respect the polarization appears as a purely induced effect essentially due to DM interactions between spins in different directions in space. This conclusion, which derives from the improper coupling relating $P_y$ to the order-parameter components (Eq. (\ref{eq:F6})), differs from the interpretation, based on DFT calculations, that spin canting and polarization mutually stabilize each other,\cite{ref:19} or from the weak frustration model\cite{ref:22} proposed for explaining the high transition temperature to the multiferroic phase.

\end{document}